\documentclass[10pt, conference, letterpaper]{IEEEtran}
\usepackage{amsmath,amssymb,amsfonts,amsthm}
\usepackage{graphicx}
\usepackage{multicol}
\usepackage[compress]{cite}
\usepackage{dblfloatfix} 
\usepackage[export]{adjustbox}
\def\BibTeX{{\rm B\kern-.05em{\sc i\kern-.025em b}\kern-.08em
    T\kern-.1667em\lower.7ex\hbox{E}\kern-.125emX}}
\usepackage{tabularx,comment,makecell}
\newcolumntype{Y}{>{\centering\arraybackslash}X}
\usepackage{bbm} % introduces for the indicator function \mathbbm{1}
\usepackage{booktabs}
\usepackage[acronym]{glossaries}
\usepackage[ruled,linesnumbered,noend,algo2e]{algorithm2e}
\let\oldnl\nl% Store \nl in \oldnl
\newcommand{\nonl}{\renewcommand{\nl}{\let\nl\oldnl}}% Remove line number for one line
\usepackage{tikz}
\usepackage{algorithm}
\usepackage[noend]{algpseudocode}
\usepackage{xspace}
\usepackage{paralist}
\usepackage{url}
\usepackage{tabularx,comment,makecell}
\usepackage{dsfont}
\usepackage{xcolor}
\usepackage{epigraph}
\usepackage{dblfloatfix}

\newcommand{\Sc}{\mathcal{S}}
\newcommand{\Cc}{\mathcal{C}}
\newcommand{\Fc}{\mathcal{F}}
\newcommand{\Kc}{\mathcal{K}}
\newcommand{\Pc}{\mathcal{P}}
\newcommand{\Vc}{\mathcal{V}}
\newcommand{\Xc}{\mathcal{X}}

\DeclareMathOperator*{\argmax}{arg\,max}

\newtheorem{prop}{Proposition}

\usepackage[many]{tcolorbox}

\definecolor{titlebg}{RGB}{100,22,72}
\definecolor{introbg}{RGB}{0,128,128}

\newtcolorbox{usecase}[1][]{
  breakable,
  enhanced,
  arc=0pt,
  outer arc=0pt,
  colframe=titlebg,
  colback=titlebg!05,
  overlay unbroken and first={
    \node[
      draw=titlebg,
      fill=titlebg,
      rotate=0,
      anchor=north west,
      text=white,
      font=\bfseries
    ]
    at (frame.north west)  
    {#1};
  }
}

\newtcolorbox{mission}[1][]{
  breakable,
  enhanced,
  arc=0pt,
  outer arc=0pt,
  colframe=introbg,
  colback=introbg!05,
  overlay unbroken and first={
    \node[
      draw=introbg,
      fill=introbg,
      rotate=0,
      anchor=north west,
      text=white,
      font=\bfseries
    ]
    at (frame.north west)  
    {#1};
  }
}

\usepackage{dblfloatfix}

\begin{document}

\title{OREO: O-RAN intElligence Orchestration\\ of xApp-based network services}

\author{F.~Mungari$^{1,2}$, C.~Puligheddu$^{1}$,  A.~Garcia-Saavedra$^{3}$, C.~F.~Chiasserini$^{1,2,4}$\\
1: Politecnico di Torino, Italy -- 2: CNIT, Italy -- 3:  NEC Labs Europe, Germany\\ 4: Chalmers University of Technology, Sweden 
\vspace{-0.3 cm}
}

\maketitle

\begin{tikzpicture}[remember picture, overlay]
  \node[draw,minimum width=4in, text width=18cm, text=black,font=\footnotesize] at ([yshift=-1cm]current page.north)
  {This paper has been accepted for publication on IEEE International Conference on Computer Communications (INFOCOM) 2024 Main conference. The final version published by IEEE is F. Mungari, Puligheddu, C., Garcia-Saavedra, A., Chiasserini, C. F., "OREO: O-RAN intElligence Orchestration of xApp-based network services," IEEE INFOCOM 2024- IEEE Conference on Computer Communications, Vancouver, BC, Canada, 2024};
\end{tikzpicture}

\begin{tikzpicture}[remember picture, overlay]
  \node[draw,minimum width=4in, text width=18cm, text=black,font=\footnotesize] at ([yshift=1.5cm]current page.south)
  {© 20XX IEEE.  Personal use of this material is permitted.  Permission from IEEE must be obtained for all other uses, in any current or future media, including reprinting/republishing this material for advertising or promotional purposes, creating new collective works, for resale or redistribution to servers or lists, or reuse of any copyrighted component of this work in other works.};
\end{tikzpicture}

\begin{abstract}
The Open Radio Access Network (O-RAN) architecture aims to support a plethora of network services, such as beam management and network slicing, through the use of third-party applications called xApps. To  efficiently provide network services at the radio interface,  it is thus essential that the deployment of the xApps is carefully orchestrated.  In this paper, we introduce OREO, an O-RAN xApp orchestrator, designed to maximize the offered services. OREO's key idea is that services can share xApps whenever they  correspond to semantically equivalent functions, and the xApp  output is of  sufficient quality  to fulfill the service requirements. By leveraging a multi-layer graph model that captures all the system components, from services to  xApps, OREO implements an algorithmic solution that selects the best service configuration, maximizes the number of shared xApps, and efficiently  and dynamically allocates resources to them. Numerical results as well as experimental tests performed using our proof-of-concept implementation, demonstrate that OREO closely matches the optimum, obtained by solving an NP-hard problem. Further, it outperforms the state of the art, deploying up to 35\% more services with an average of 30\% fewer xApps and a similar  reduction in the resource consumption. 
\end{abstract}

\section{Introduction}
As mobile networks continue to gain momentum, it has become essential to accommodate a growing number of traffic classes and 
demanding services \cite{6g_network}. The 
O-RAN architecture \cite{ORANWG1_Arch}, promoted by the O-RAN Alliance, addresses such a need by  building upon virtualization and openness principles  
\cite{understanding_oran, oran_disrupting_vran_ecosystem}. 
A key feature of the O-RAN architecture is network automation, 
enabled by \textit{xApps} and \textit{rApps}, 
third-party applications running in Radio Intelligent Controllers (RICs) and  operating, respectively,  below and above the 1-second  timescale. 
In this context, the integration of  Machine Learning (ML)  
plays a crucial role, since     
ML-based xApps can serve as the foundation of the RAN intelligence for closed-loop and  
agile network control~\cite{ORANWG1_UseCases}.  
However, 
as the use of ML takes a substantial toll on the system resources, it is imperative to devise novel 
orchestration policies that effectively minimize the xApps computational footprint within the O-RAN architecture. 
{\color{black}This  is especially important when resources 
are scarce and shared among multiple tenants. Reducing resource utilization not only decreases the operational expenditure of RANs, representing 40\% of the total costs in cellular network~\cite{RAN_related_costs}, but it also contributes to lower  energy consumption. 
}

\textbf{Existing research gap.} The design of an efficient policy for the orchestration of RAN 
intelligence in O-RAN platforms is still an open challenge. 
The state of the art, discussed in details in Sec.~\ref{sec:related_work}, either fails to acknowledge the complexity of the 
RAN intelligence management in  O-RAN, or only considers monolithic services and xApps with oversized resource allocations, 
resulting in sub-optimal decisions.
To  fill this gap, we propose a novel O-RAN intElligence Orchestration (OREO) 
framework that, given the Mobile Network Operators' (MNO) demand for network services, 
 identifies {\em the  set of xApps to 
deploy as well as their operational configuration}, while meeting the specific service requirements and keeping resource expenditure at the minimum.

\textbf{OREO distinctive features.} 
Our study pioneers the integration of the Network Function Virtualization (NFV) paradigm 
into the O-RAN architecture, recognizing that RAN services, such as beam allocation and handover prediction, 
can be built by interconnecting elementary RAN functions, like load forecaster and traffic classificator.
Importantly, our approach facilitates the sharing of common functions across services and efficient tuning of resources allocated to contributing xApps.  
As detailed in Sec.~\ref{subsec:oreo_features}, we consider that functions with the same semantic can be deployed as xApps at distinct complexity levels, leading to different trade-offs among function output quality, resource consumption, and execution speed.

\textbf{Technical challenges.}
Jointly managing the orchestration of RAN services and associated resources opens the door to an enhanced orchestration policy.
However, to grasp such an opportunity, an O-RAN service orchestrator has to address multiple  challenges: 
($i$) it has to identify the most convenient service configuration, i.e., the xApps, from those available in the catalog of the near-real-time (near-RT) RIC, 
that have the right semantic to implement a service matching the desired functional requirements;  
($ii$) the output quality, hence the level of complexity, of each  xApp must adhere to the service quality requirements;
($iii$) the computational (e.g., CPU, GPU) and memory  (e.g., RAM, disk) resources allocated to the  xApps must be sufficient for fulfilling the service latency requirements, while not exceeding the available resource budget.

\textbf{Summary of novel contributions.}

\textbullet~By  exploiting the NFV paradigm, we  look at O-RAN services not as monolithic, rigid entities, but rather as 
  sets of interconnected elementary functions that can be implemented as O-RAN xApps with distinct levels of complexity 
  (i.e., yielding different levels of output quality). This approach dramatically increases the flexibility 
  in deploying RAN services. 

\textbullet~By developing a model that captures all relevant aspects, we formulate the (NP-hard) xApp Deployment and Sharing (xDeSh) problem that maximizes the {\color{black} MNO profit generated by} the offered services while accounting for the service and system constraints.
 
\textbullet~We design an intelligent  orchestrator -- named OREO -- and integrate it into the O-RAN architecture.  
In view of the complexity of the xDeSh problem, the OREO engine implements a heuristic solution to xDeSh that: 
($i$) selects the xApps providing the functions that are semantically  necessary for the service deployment, 
($ii$) at the level of complexity  that best trades off  output quality with resource expenditure.
To our knowledge, OREO is the first to leverage the NFV paradigm for an efficient and flexible deployment of O-RAN xApps.

\textbullet~We evaluate OREO through an extensive numerical analysis  
and  experimental results obtained with our  proof-of-concept testbed, integrated in a O-RAN platform and running real-world RAN services. OREO can support a number of services close to the optimum, and, compared to the state of the art, it enables the co-existence of more services  (16.3\% more on average and up to 25.8\%), while reducing resource expenditure (by 28.8\% less on average and up to 35\%).

\section{The OREO Framework}
\label{sec:OREO}

This section presents OREO,  outlining first its purpose  and the distinctive features of its engine (Sec.~\ref{subsec:oreo_features}), and then  the rationale behind its design and its integration within the O-RAN architecture (Sec.~\ref{subsec:system_architecture}).

\subsection{OREO driving purpose and distinctive features}\label{subsec:oreo_features}

The success of next-generation mobile networks greatly hinges upon the quality of RAN intelligence and, hence, upon the performance of its orchestration framework, which is responsible for deploying RAN management services~\cite{orchestration_requirements}. 
Our orchestrator, OREO, acts upon a set of service requests by the MNOs in an O-RAN platform. Given such requests, OREO selects the xApp(s) required to deploy the services in the near-RT RIC, and the  specific xApps configuration that lets a service meet its performance requirements  while matching the resource availability in the platform. 
Importantly, despite this work focuses on xApps, the design of the OREO framework is not limited to near-RT services and can be extended to non-RT RAN functions to optimize the instantiation and management of rApps within the non-RT RIC.
OREO's orchestration decisions are made by its core component, the OREO engine, which greatly differs from state-of-the-art O-RAN orchestrators such as the pioneering work in~\cite{orchestran}. 

A key differentiating principle that drives the design of the OREO engine consists in conceiving O-RAN services as sets of interconnected functions rather than monolithic entities. This approach capitalizes on the known benefits of the NFV paradigm, such as enhanced flexibility, scalability, and cost-effectiveness~\cite{service_function_chaining}. By recognizing that O-RAN services can be built by interconnecting elementary RAN management functions, OREO leverages xApps as fundamental building blocks to efficiently offer such services.  
More specifically, 
\begin{itemize}
    \item {\em Service composition:} Each network service request fed by an MNO to the OREO engine is associated with a minimum service quality and a maximum latency requirement. 
    To meet performance targets, services can be deployed  using different configurations: each configuration corresponds to a different set of functions, with each function implementing a certain task. For instance, enabling network slicing may involve a single function implementing a reinforcement learning (RL)-based policy \cite{xApp_ColO-RAN}, or combining it with a traffic predictor to enhance system response to traffic condition changes.
    \item {\em Implementing service functions through xApps:}  Additionally, each function can be implemented through multiple xApps, each executing semantically the same function, but instantiated at a different operating point, hereinafter also referred to as \emph{complexity factor}. Importantly, complexity factors provide different trade-offs between the output quality and processing latency of the function offered by the xApp and the computational resources necessary to run that xApp. 
\end{itemize}

Thus, based on the above concepts, in OREO the quality and latency incurred by a service depend upon:
\begin{itemize}
    \item The specific configuration (set of functions) that is selected to enable the service;
    \item The specific xApps (hence levels of complexity) that are chosen to implement the functions in the selected configuration.
\end{itemize}

The OREO engine identifies the service configuration {\em and} the corresponding xApps  in such a way that it can best suit the service requirements. 
Furthermore, 
\begin{itemize}
    \item Whenever multiple services require the same function, {\em OREO allows such services to share the xApp that implements the semantics of that function}, if the xApp complexity level meets the  requirements of the services;  
\item As  the service latency targets can be fulfilled by properly setting the resources allocated to the shared xApps, {\em OREO scales the resources assigned to an xApp according to the overall load imposed by the corresponding services}, as well as the available resource budget. Importantly, in so doing, OREO avoids resource over-provisioning, as opposed to relying upon a fixed amount of resources allocated to xApps as in state-of-the-art solutions~\cite{orchestran}.
\end{itemize}

\subsection{OREO system architecture}
\label{subsec:system_architecture}

\begin{figure}[tbp]
    \centering
    \includegraphics[width=0.35\textwidth]{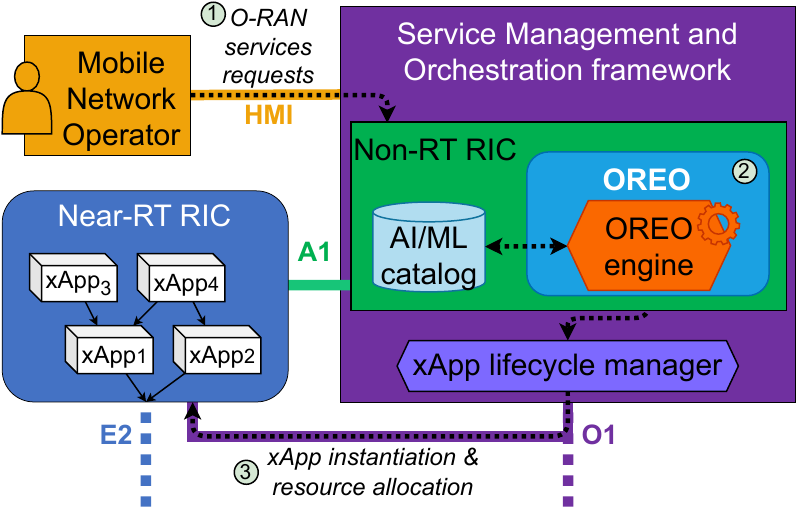}
    \caption{OREO design and integration in the O-RAN architecture. The workflow (dashed black line) is as follows: ($i$) the MNO submits service requests via the Human-Machine interface (HMI); ($ii$) OREO processes such requests and, with the support of the xApp lifecycle manager, instructs the near-RT RIC via the O1 interface about which xApps to deploy.}
\label{fig:oreo_system_design}
\vspace{-4mm}
\end{figure}

The OREO framework, illustrated in Fig.~\ref{fig:oreo_system_design}, 
is designed to be integrated into the O-RAN Service Management and
Orchestration (SMO), which is responsible for managing and orchestrating all control and monitoring procedures of the RAN components via the O1 interface. In particular, OREO operates within the non-RT RIC, which supports the execution of third-party applications known as rApps and, through the A1 interface, enables closed-loop control of the RAN. 

By accessing the Human-Machine interface (HMI), OREO receives management intents from an MNO, which submits requests for a set of services, specifying  
the maximum tolerable delay and minimum quality for each service {\color{black}(Step~1)}. 
As mentioned in the previous section, the OREO engine calculates the deployment of xApps satisfying the service requests {\color{black}(Step~2)}. xApps are indeed third-party applications that implement customized logic to drive the RAN efficiently and run within the near-RT RIC, i.e., the central control and optimization unit of the RAN operating on a sub-second time scale.
{\color{black} The selected xApps are deployed within the the near-RT RIC using management services, such as the xApp lifecycle manager provided by the SMO through the O1 interface (Step~3).}
Through the E2 interface and open APIs, the near-RT RIC interacts with the RAN centralized and distributed units (O-CUs and O-DUs, respectively), collecting RAN performance metrics and providing control actions. 

As detailed in Sec.\,\ref{sec:performance_eval}, we have implemented all OREO components and integrated them in the 
O-RAN architecture, leveraging an O-Cloud platform \cite{o-cloud} that hosts the SMO, the non-RT RIC, and the near-RT RIC. So doing, we have developed a proof-of-concept testbed used to measure the performance of the proposed solution in real-world settings.  

\section{xDeSh: xApp Deployment and Sharing}
\label{sec:xdesh}
We now present the system  model (Sec.~\ref{subsec:system_model}) and  the  xApp Deployment and Sharing (xDeSh) problem (Sec.~\ref{subsec:problem_formulation}).

\begin{figure}[tbp]
    \centering
    \includegraphics[width=0.35\textwidth]{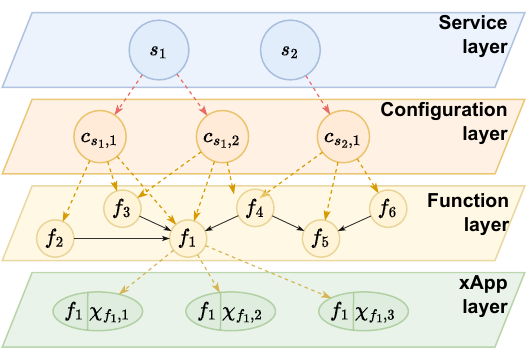}
    \caption{Graph-based representation of the system under study and relation between its main components. For clarity, the xApp layer only includes the xApps implementing function $f_1$.}
    \label{fig:system_model_graph}
\vspace{-4mm}
\end{figure}

\subsection{System model \label{subsec:system_model}}

 Fig.~\ref{fig:system_model_graph} depicts the  system components, which we further detail in the following; Tab.~\ref{tab:notation} summarizes the used notations.

\begin{table}[tbp]
\caption{Notations\label{tab:notation}}
\vspace{-1mm}
\footnotesize
    \footnotesize 
    \begin{tabularx}{\columnwidth}{|c|X|}
    \hline
    \multicolumn{2}{|c|}{\bf Parameters}\\
    \hline
    \hline
    {\bf Symbol} & {\bf Description}\\
    \hline
    $s {\in} \Sc$ & RAN service \\
    \hline
    $T_s$ ($Q_s$) & Target latency of service $s$ under configuration $c_s$\\
    \hline
    $p_s$ & Priority of service $s$ \\
    \hline
    $c_s {\in} \Cc_s$ & Service $s$ configuration \\
    \hline
    $\Vc_{c_s}$ & Set of nodes of the service configuration graph $c_s$ \\ 
    % \hline
    % $\Ec_{c_s}$ & Set of edges of the service configuration function graph $c_s$ \\ 
    \hline
    $f{\in} \Fc$ & RAN function\\
    \hline
    $\chi_f {\in} \Xc_f$ & Complexity factor of function $f$\\
    \hline
    $f_\chi$ & xApp implementing function $f$ with complexity $\chi$ \\
    \hline
    $f_\chi^{(j)}$ & $j$-th instance of xApp $f_\chi$ \\
    \hline
    $\mu_{f_\chi^{(j)},\text{mem\,(disk)}}$ & Memory (disk) requirement of xApp $f_\chi^{(j)}$\\
    \hline
    $\lambda_{\Pc(f_{\chi}^{(j)})}$ & Input data rate of $f_\chi^{(j)}$ if shared among $s {\in} \Pc(f_{\chi}^{(j)})$ \\
    \hline
    $\theta_{f_\chi}$ & Amount of input data processed by ${f_\chi}$ in a CPU cycle \\
    \hline
    $\Kc$ & Set of resource types \\
    \hline
    $\mathbf{B}$ & Vector of resource budgets of the different types\\
    \hline
    $q_{c_s,f_{\chi}}$ & Quality of xApp $f_\chi$ \\
    \hline
    $l_{f_\chi^{(j)}}$ & Processing latency of the j-th instance of xApp $f_\chi$ \\
    \hline
    $\tau_{c_s}$ & Latency of service $s$  \\
    \hline
    \end{tabularx}

    \vspace{1mm}
    
    \begin{tabularx}{\columnwidth}{|c|X|}
    %\hline
    \hline
    \multicolumn{2}{|c|}{\bf Decision variables}\\
    \hline
    \hline
    {\bf Symbol} & {\bf Description}\\
    \hline
    $z_{c_s}$ & Binary variable for service configuration  $c_s$ selection \\
    \hline
    $v_{ c_s, f_\chi^{(j)} }$ &  Binary variable indicating if $f_\chi^{(j)}$ is used in configuration $c_s$ \\
    \hline
    $\boldsymbol{\rho}_{f_\chi^{(j)}}$ & Resource allocation for the j-th instance of xApp $f_\chi$\\
    \hline
    \end{tabularx}
\vspace{-4mm}
\end{table}

\textbf{\textbullet~{Services and service configurations.}}
We focus on decision-making services managed by the near-RT RIC. 
%such as slice resource allocation and traffic steering services. 
Each service $s$ is characterized by: ($i$) a target latency $T_s$ specifying the maximum acceptable delay to output a decision since a service request arrives; and ($ii$) a target decision quality $Q_s$, e.g., the minimum required accuracy for a traffic classification task. 
Further, a service is assigned a priority level $p_s$, which {\color{black} depends on the revenue generated for the MNO and is used to} determine which services should be dropped in case of insufficient resource availability.
A service can be provided using different configurations,  $c_s \in \Cc_s$, i.e., sets of interconnected elementary functions.  
Each configuration is associated with a level of quality and resource demand, determined by the set of functions appearing in the configuration; thus,  properly selecting $c_s$ makes it  possible to  trade off  the   performance of a service with its deployment and running cost. 

\textbf{\textbullet~{Functions and xApps.}}
A function $f {\in} \Fc$ represents a low-level operation and serves as the fundamental building block of one or more services. Examples of functions include traffic forecasting and traffic classification. Functions may process ($i$) metrics collected by the RAN elements (O-DU, O-CU, etc.) and shared with the near-RT RIC via the E2 interface; and/or ($ii$) information provided by other functions. A service configuration can then be modeled as a directed graph whose vertices ($\Vc_{c_s}$) and edges represent, respectively, the functions composing the configuration and the dependency relations between them. Specifically, an edge exists from function $f^\prime$ to function $f$ whenever the execution of $f$ requires the output of $f^\prime$.
Each function can be implemented with a different \emph{complexity factor}, $\chi_f{\in} \Xc_f$.
For instance, a traffic classification function can be provided by different ML models, each offering a different accuracy-resource demand trade-off.

A specific function with a given complexity factor defines an xApp, which is thus indicated as $f_\chi=(f,\chi_f)$. 
Let  $\Pc(f_{\chi})$ be the set of service configurations that include xApp $f_{\chi}$, and  $\lambda_{\Pc(f_{\chi})}$ be the rate at which data is fed to, and needs to be processed by, the xApp per unit of time.
We remark that,  multiple instances (i.e., replicas) of a given xApp can be implemented and, hence, coexist in the system;  we then denote the $j$-th instance of ${f_{\chi}}$ with ${f^{(j)}_{\chi}}$.

\textbf{\textbullet~{Near-RT RIC resources.}}~The O-Cloud can provide the near-RT RIC with computing and storage resources (e.g., CPU, GPU, memory) to run xApps. We denote with $\Kc{=} \{ 1, \, \ldots, \, K \}$ the set of available resource types, and with 
$\boldsymbol{B} = [ B_1, \, \ldots , \, B_K ]$ the vector collecting, for each  type, the available resource budget. 
For simplicity and without loss of generality, in the following we focus on CPU, memory, and disk storage. 
Thus, we let  
$\boldsymbol{\rho}_{f_\chi^{(j)}} {=} [\rho_{f_\chi^{(j)},1},\allowbreak \, \ldots,\allowbreak \, \rho_{f_\chi^{(j)},K}]$, with $\rho_{f_\chi^{(j)},k}{\leq} \boldsymbol{B_k}$ denote the amount of resource of type $k$ reserved for the xApp instance $f_\chi^{(j)}$,
{\color{black} 
and the corresponding memory and disk requirements with, respectively, $\mu_{f_\chi^{(j)},\text{mem}}$ and $\mu_{f_\chi^{(j)},\text{disk}}$}. 

\textbf{\textbullet~{xApp and service quality.}} 
Let $q_{c_s,f_{\chi}}$ be the quality score obtained by the xApp $f_{\chi}$ associated with service  configuration $c_s$. Common quality metrics include prediction and classification accuracy, regression error, and expected reward.
The quality score $q_{c_s,f_{\chi}}$ depends on the quality of the input data, which, in turn, depends on the complexity level associated with the function $f^\prime$ preceding $f$ in the configuration graph.
Accordingly, the quality metric for a service $s$ implemented under configuration $c_s$, denoted as $q_{c_s}$, is equal to the  quality of the last xApp's output in the configuration graph.

\textbf{\textbullet~{xApp and service latency.}} 
Given the considered resource types, {\color{black}as long as memory and disk requirements ($\mu_{f_\chi^{(j)},\text{mem}}$ and $\mu_{f_\chi^{(j)},\text{disk}}$) are satisfied}, only the CPU allocation has an impact on the xApps processing latency, denoted by $l_{f_\chi^{(j)}}
% (\rho_{f_\chi^{(j)},\text{cpu}})
$ for the $j$-th instance of xApp $f_\chi$. 
Drawing on the existing works~\cite{Cohen,mm1_1,mm1_2}, we can model a function, $f_\chi^{(j)}$, that is shared among the service configurations in ${\Pc}(f_\chi^{(j)})$, as an M/M/1 queue. We can then write the corresponding average processing latency as:
%{\color{black}REV.1-3 and REV.6-2:No dependency on $\rho$ explicated
\begin{equation*}
    \label{eq:xapp_lat}
    l_{f_\chi^{(j)}}
    = 
    ( \rho_{f_\chi^{(j)},\text{cpu}} {\theta_{f_\chi^{(j)}}}- \lambda_{\Pc(f_{\chi}^{(j)})})^{-1}
\end{equation*}
%}
where $\rho_{f_\chi^{(j)},\text{cpu}}$ is expressed as CPU cycles per second, ${\theta_{f_\chi^{(j)}}}$ represents the xApp complexity and expresses the amount of input data processed by the xApp in a CPU cycle, {\color{black} and $\lambda_{\Pc(f_{\chi}^{(j)})}$ specifies the xApp load when shared among ${\Pc(f_{\chi}^{(j)})}$ service configurations.}
Also, let  $\tau_{c_s}$ be the latency of service $s$ when implemented with configuration $c_s$.
Defining a path $\pi_{c_s}$ on the graph of configuration $c_s$ as a set of edges connecting an input function with an output function in $c_s$, $\tau_{c_s}$ is the latency associated with the most time-consuming path in the graph.
{\color{black} 
The latency for collecting data is indeed deemed negligible as the near-RT RIC periodically exposes data to the xApps.}
The path latency  depends on the complexity factor and resource allocation of each of the functions composing the path, 
\begin{equation}
    \label{eq:service_latency}
    \tau_{c_s}
    =
    \argmax_{\{ \pi_{c_s} \}}
    \sum_{f \in \pi_{c_s}}
    l_{f_{j,\chi_{f_j}}} 
    \,.
\end{equation}
%}

\subsection{xDeSh problem formulation 
\label{subsec:problem_formulation}}
Given the above model, we now introduce the xDeSh  optimization problem, along with some additional system variables and parameters defining the current state of the system. Further, we prove that  the xDeSh problem is NP-hard.

\textbf{\textbullet~{Service configuration selection.}}
Let $\Sc$ be the set including both the existing, and still to be kept, services and the new services to be deployed.
For each service $s{\in}\Sc$, the OREO orchestrator identifies the most suitable configuration $c_s$ to be used.
We denote with $z_{c_s}$ the binary decision variable taking 1 if configuration $c_s$ is selected for service $s$. 
Notice that: ($i$) it may happen that none of the possible  configurations of a service $s$ can be deployed, due to insufficient resources to guarantee the minimum required service performance; ($ii$) at most one configuration per service can be selected. That is, the following constraint must hold:
\begin{equation}
    \label{eq:constraint_servconf}
    \sum_{c_s \in \Cc_s}{ z_{c_s}} \le 1, \, \forall s \in \Sc \,.
\end{equation}

\textbf{\textbullet~{Selection of xApps to implement and share.}}
Whenever an xApp $f_\chi$ is required by more than one service, the orchestrator has to determine whether to let such services share the same instance $f_\chi^{(j)}$, or to implement multiple instances thereof. 
We thus introduce the binary decision variable $v_{ c_s, f_\chi^{(j)} }$, to indicate whether $f_\chi^{(j)}$ is used by service configuration $c_s$ or not.
Clearly, all the functions required by a selected service configuration must be implemented.
Moreover, neglecting the possibility of scaling out xApps,  a service configuration  cannot use more  than one instance of a given xApp. The above requirements translate in the following constraint:
\begin{equation}
    \label{eq:constraint_f_in_cs}
    \sum_{\chi \in \Xc_f} \sum_j {v_{ c_s, f^{(j)}_{\chi}}} = z_{c_s}
    \textnormal{, }
    \forall s{\in}\Sc,
    \forall c_s{\in}\Cc_s,
    \forall f{\in}\Vc_{c_s} \,.
\end{equation}
Similarly, an xApp implementing function $f$ cannot be associated with a service configuration that does not include $f$:
\begin{equation}
    \label{eq:constraint_f_notin_cs}
    \sum_{\chi \in \Xc_f} \sum_j {v_{ c_s, f^{(j)}_{\chi}}} = 0
    \textnormal{, }
    \forall s{\in}\Sc,
     c_s{\in}\Cc_s,
     f{\notin}\Vc_{c_s}\,.
\end{equation}

{\color{black} 
Ultimately, the orchestrator allocates memory and disk resources for deploying the necessary xApps, while adhering to the following constraint:
\begin{equation}
    \label{eq:xApp_storage_res}
    \rho_{f_\chi^{(j)},k}
    {\ge}
    \mu_{f_\chi^{(j)},k} \mathbbm{1}_{ [ \exists c_s{\in}\Cc_s | {v_{ c_s, f^{(j)}_{\chi}}} {=} 1] }
    \textnormal{, }
    \forall
    k {\in} \{\text{disk,\,mem}\}
\end{equation}
where the indicator function $\mathbbm{1}$ equals one when the subscripted condition holds, indicating that xApp $f_\chi^{(j)}$ must be deployed.
}

\textbf{\textbullet~{Meeting service requirements.}}
OREO has to  select a service configuration (i.e., the functions that implement the service) and the corresponding resources in such a way that the  quality and latency targets are satisfied. That is, for any $s {\in} \Sc$ and $c_s {\in} \Cc_s$,
\begin{eqnarray}
    \label{eq:service_quality_constraint}
&&    q_{c_s}
    \ge    Q_s  z_{c_s}  \\
&&    \tau_{c_s}
    z_{c_s}   {\le}  T_s \,. 
    \label{eq:service_latency_constraint}
\end{eqnarray}

\textbf{\textbullet~{Complying with the resource budget.}}
We also need conventional capacity constraints, i.e., the near-RT RIC resource budget $\boldsymbol{B}$ must not be exceeded:
\begin{equation}
    \label{eq:constraint_budget}
    \sum_{f {\in} \Fc} \sum_{\chi {\in} \Xc} \sum_{j}
    {\rho_{f^{(j)}_{\chi},k}} \le B_k,\; \forall k {\in} \Kc \,.
\end{equation}

\textbf{\textbullet~{Near-RT RIC's settings.}}
Let  $ {\hat{\Sc}} = \{ \hat{s} \}$ denote the set of services that are already deployed, and $c_{\hat{s}}$ capture their service configuration. Accordingly, the binary parameter $\hat{z}_{c_{\hat{s}}}$ takes 1 if  configuration  $c_{\hat{s}}$ of service $\hat{s}$ is implemented, and 0 otherwise. 
Now, given an xApp instance $f_\chi^{(j)}$, we let $\hat{v}_{ c_{\hat{s}}, f_\chi^{(j)} }$ indicate whether service configuration $c_{\hat{s}}$ is using $f_\chi^{(j)}$, and $\hat{\boldsymbol{\rho}}_{f_\chi^{(j)}}$ denote  the  vector indicating its current resource allocation.  

\textbf{\textbullet~{No service disruption.}}
It is critical to account for the cost incurred by the system whenever OREO determines a new configuration for an existing service $s$. Indeed, due to the need to ensure  continuity for a service $s{\in} \hat{\Sc}{\cap} \Sc$, 
both the xApps required by $ \{ c_{\hat{s}} \, | \, \hat{z}_{c_{\hat{s}}} \}_{\hat{s} \in \hat{\Sc} \cap \Sc} $ and by $ \{ c_s \, | \, z_{c_s} \}_{s \in \hat{\Sc} \cap \Sc}$ have to co-exist before ($i$) turning off the relative currently implemented, but no longer required, xApps, and ($ii$)  instantiating the remaining functions required by the residual services in $ \Sc$.
We then define $\Fc_{1}$ as the set of xApps required by the existing services that  should not be deactivated:
\begin{gather*}
    \Fc_{1} = 
    \{ f^{(j)}_\chi \, | \,
    \sum_{c_{\hat{s}} {\in} \Cc_{\hat{s}}}
    \hat{v}_{ c_{\hat{s}}, f_\chi^{(j)} } = 1 
    \}_{ f {\in} \Fc, \, \chi_f {\in} \Xc_f, \, j , \, \hat{s} {\in} \hat{\Sc} {\cap} \Sc} \,.
\end{gather*} 
Similarly, we define $\Fc_{2}$ as the set of xApps required in the last defined near-RT RIC's setting for the services whose operation must not be disrupted.
% \begin{gather*}
%     \Fc_{2} = 
%     \{ f^{(j)}_\chi \, | \,
%     \sum_{c_{{s}} {\in} \Cc_s}
%     \hat{v}_{ c_s, f_\chi^{(j)} } = 1 
%     \}_{ f {\in} \Fc, \, \chi_f {\in} \Xc_f, \, j , \, s {\in} \hat{\Sc} {\cap} \Sc} \,.
% \end{gather*}
Then,  we must have:
\begin{equation}
    \label{eq:constraint_service_disruption}
    \sum_{f^{(j)}_{\chi} {\in} \Fc_1 \backslash \Fc_2 }{ \hat{\rho}_{f^{(j)}_\chi,k}}
    +
    \sum_{f^{(j)}_{\chi} {\in} \Fc_2 }
    {\rho_{f^{(j)}_{\chi},k}}
    {\le} B_k
    \textnormal{, } \forall k {\in} \Kc
\end{equation}
where $ \Fc_1 \backslash \Fc_2 {=} \{ f^{(j)}_{\chi} \, | \, f^{(j)}_{\chi} {\in} \Fc_1 {\land} f^{(j)}_{\chi} {\notin} \Fc_2 \}$.

\textbf{\textbullet~{Objective function.}}
The xDeSh problem  defines an xApp selection and resource allocation policy that ($i$) maximizes the number of offered services {\color{black}based on their priority levels}, and ($ii$) minimizes the near-RT RIC resource consumption, i.e.,
\begin{equation}
    \label{eq:obj_function}
    \Psi(\boldsymbol{z}, \boldsymbol{v} , \boldsymbol{\rho})
    {=} 
    {\sum_{s {\in} \Sc}\sum_{c_s {\in} \Cc_s} z_{c_s} \, p_s}
    {-}
    \frac{1}{K}
    \sum_{f {\in} \Fc}\sum_{\chi {\in} \Xc_f}\sum_{j}\sum_{k {\le} K}{ \frac{ \rho_{f^{(j)}_{\chi},k}}{B_k}} \nonumber
\end{equation}
where the decision variables (see Tab.~\ref{tab:notation}) have been vectorized,
{\color{black}and the $1/K$ factor prevents service rejection for the sake of resource savings.}
The xDeSh problem can then be formulated as:

\begin{usecase}[xApp Deployment and Sharing (xDeSh) Problem]
\begin{equation}
\begin{aligned}
\max_{
    \boldsymbol{z}, \boldsymbol{v},\boldsymbol{\rho}
    }
    \enspace & \Psi(  \boldsymbol{z}, \boldsymbol{v}, \boldsymbol{\rho} ) \\
\textrm{s.t.} \enspace & 
    (\ref{eq:constraint_servconf}),
    (\ref{eq:constraint_f_in_cs}),
    (\ref{eq:constraint_f_notin_cs}),
    {\color{black}(\ref{eq:xApp_storage_res}),}
    (\ref{eq:service_quality_constraint}),
    (\ref{eq:service_latency_constraint}),
    (\ref{eq:constraint_budget}),
%    (\ref{eq:unimplemented_xApps}),
    (\ref{eq:constraint_service_disruption})\\
& z_{c_s} \in \{0,1\} \enspace \forall s \in \Sc, c_s \in \Cc_s \\
& v_{ c_s, f^{(j)}_{\chi_f}} \in \{0,1\} \enspace \forall c_s \in \Cc_s, f \in \Fc, \chi_f \in \Xc_f, j \\
& \rho_{f^{(j)}_{\chi},k} \in [0,B_k] \enspace \forall k \in \Kc, f \in \Fc, \chi_f \in \Xc_f, j\nonumber
\end{aligned}
\end{equation}
\end{usecase}

\begin{prop}
\label{theorem:np-hard}
The xDeSh problem is NP-hard.
\end{prop}

\begin{proof}
%\textit{Proof.}
The full proof is omitted for brevity. The thesis in proved by showing that any instance of the well-known NP-hard multi-commodity facility location problem (FLP) can be reduced  to an instance of the xDeSh problem. 
As the  reduction can be obtained in polynomial time, the thesis follows.  
\end{proof}

\section{Solving the xDeSh Problem}
\label{sec:heuristic}

Motivated by Proposition~\ref{theorem:np-hard} above, we propose an efficient heuristic solution,
which is outlined in Sec.~\ref{subsec:iterative_solution}.  
Then, we describe in detail each of its building blocks, namely, a Lagrangian relaxation  and a decoupling method
(Sec.~\ref{subsec:lagrangian_problem}), a feasibility testing algorithm  (Sec.~\ref{subsec:feasible_algorithm}), 
and, finally, a subgradient method (Sec.~\ref{subsec:subgradientmethod}). 
We remark that our solution algorithm resides in the OREO engine, introduced in Sec.\,\ref{sec:OREO}.

\begin{figure}[tbp]
    \centering
    \includegraphics[width=0.45\textwidth]{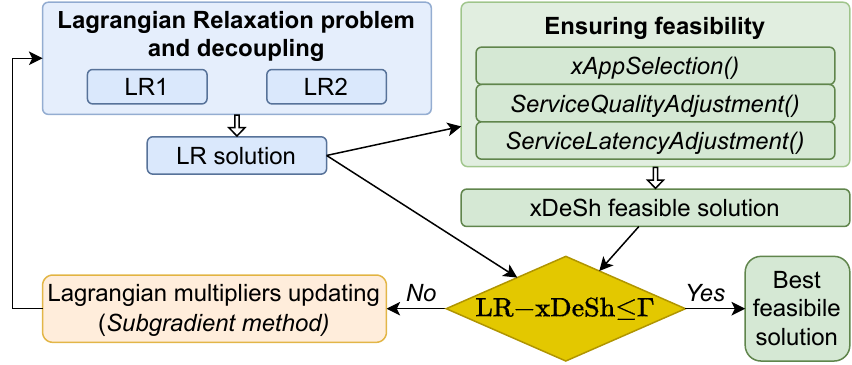}
    \caption{The xDeSh problem is solved with an iterative algorithm that alternates the Lagrangian relaxation and 
    the subgradient method until the set stopping criterion is met.}
    \label{fig:heuristic_scheme}
\vspace{-4mm} \end{figure}

\subsection{Overview of the algorithmic solution}  \label{subsec:iterative_solution}
In the proof of Prop.~\ref{theorem:np-hard}, we underlined the similarity between the xDeSh problem and the FLP. Then, inspired by existing efficient FLP solvers~\cite{lagr_heuristic_2CFLP_1, lagr_heuristic_2CFLP_2, lagr_heuristic_2UFLP}, we design our algorithmic solution adopting an iterative, 
%including both one-stage~\cite{lagr_heuristic_1UFLP} and 
two-stage approach.  
As illustrated in Fig.~\ref{fig:heuristic_scheme}, our solution framework first leverages the Lagrangian Relaxation (LR) method, 
a relaxation technique that incorporates the effect of the constraints that entail the problem's complexity into the objective function. To enforce these constraints, the method introduces penalty terms, 
known as Lagrange multipliers. However, this approach may provide a solution to the xDeSh problem that is not feasible.

To solve this issue, we combine the LR method with an algorithm capable of identifying the violated constraints and  making  adjustments to the relaxed solution. Importantly, the feasible and infeasible solutions that we get represent the lower and upper bounds on the optimal solution, respectively. 
To obtain increasingly tighter bounds, we leverage the subgradient method -- a robust technique that provides a policy for progressively updating the Lagrangian multipliers.
% until convergence is achieved. At the point of convergence, the heuristic procedure provides the best found feasible solution so far.

The above solution process is repeated until one of the three stopping criteria is met. 
The first criterion terminates the process when the LR and the obtained solutions differ by less than a given threshold, $\Delta$. Subsequently, the iterative process is stopped if the step size,  determining the size of updates to the Lagrangian multipliers through the subgradient method, drops below a designated threshold $\Gamma$. Indeed, the step size is initially set to large values to facilitate rapid updates and then halved when the iterative process fails to improve the solution for $N$ iterations, aiming to refine and stabilize the overall process.
The third stopping criterion finally sets a predefined maximum number of overall iterations, $\Lambda$. 

\subsection{Problem relaxation and decoupling} \label{subsec:lagrangian_problem}

To apply the LR to the xDeSh problem, we note that 
constraints (\ref{eq:constraint_f_in_cs}), (\ref{eq:service_quality_constraint}), and (\ref{eq:service_latency_constraint}) 
entangle the service configuration and the xApp selection subproblems. However, since the LR deals with inequalities, 
we split constraint (\ref{eq:constraint_f_in_cs}) into:
\begin{eqnarray}
    \label{eq:constraint_f_in_cs_A}
    \sum_{\chi \in \Xc_f} \sum_j {v_{ c_s, f^{(j)}_{\chi}}} {\geq} z_{c_s}
    \textnormal{, } & &
    \forall s {\in}  \Sc,
    c_s {\in}  \Cc_s,
    f {\in}  {\Vc}_{c_s}\\
    \label{eq:constraint_f_in_cs_B}
    \sum_{\chi {\in}  \Xc_f} \sum_j {v_{ c_s, f^{(j)}_{\chi}}} {\leq} 1 
    \textnormal{, } & &
    \forall s {\in} \Sc,
     c_s {\in} \Cc_s,
   f {\in} {\Vc}_{c_s} \,.
\end{eqnarray}
The two inequalities above indeed provide, respectively, a 
lower and an upper bound on the number of xApps implementing the same function for a given service configuration $c_s$, and they collapse into (\ref{eq:constraint_f_in_cs})  for the selected configuration. 
Moreover, we linearize (\ref{eq:service_latency_constraint}), which links the configuration selection with 
the relative expected latency by adopting the big-M linearization for each service $s$ implemented according to configuration $c_s$:
\begin{equation} \label{eq:service_latency_constraint_linear}
    \tau_{c_s}
    {-} T_s {\leq} M(1 {-} 
    z_{c_s} )\,.
\end{equation}

We relax constraints  (\ref{eq:constraint_f_in_cs_A}), (\ref{eq:service_quality_constraint}) and 
(\ref{eq:service_latency_constraint_linear}) by introducing, respectively, the non-negative Lagrangian penalty 
terms $\boldsymbol{\beta}{=} \{ \beta_{c_s, f} \}_{s, c_s, f} $, $\boldsymbol{\gamma} {=} \{ \gamma_{c_s} \}_{s, c_s} $, and 
$\boldsymbol{\delta}{=} \{ \delta_{c_s} \}_{s, c_s} $, which leads to the below LR formulation. 

\textbf{\textsc{xDeSh Lagrangian Relaxation problem (LR):}}
\begin{equation*}
\begin{aligned}
\max_{
    \boldsymbol{z}, 
    \boldsymbol{v}, 
    \boldsymbol{\rho} 
    }
    \enspace & \Psi_{L} ( \boldsymbol{z}, \boldsymbol{v}, \boldsymbol{\rho}, \boldsymbol{\beta}, \boldsymbol{\gamma}, \boldsymbol{\delta} ) \\
\textrm{s.t.} \enspace & 
    (\ref{eq:constraint_servconf}),
    (\ref{eq:constraint_f_notin_cs}),
    {\color{black}(\ref{eq:xApp_storage_res}),}
    (\ref{eq:constraint_budget}),
    (\ref{eq:constraint_service_disruption}),
    (\ref{eq:constraint_f_in_cs_B})\\
\end{aligned}
\end{equation*}
with the Lagrangian function $\Psi_{L}$  defined as:
\begin{equation}
\begin{cases}
    \Psi_{L}( \boldsymbol{z}, \boldsymbol{v}, \boldsymbol{\rho}, \boldsymbol{\beta}, \boldsymbol{\gamma}, \boldsymbol{\delta} ) 
    {=}  \Psi_{L,1} {+} \Psi_{L,2}&\nonumber\\
    
    \Psi_{L,1}(\boldsymbol{z}, \boldsymbol{\beta}, \boldsymbol{\gamma}, \boldsymbol{\delta}) {=} 
        \sum_{c_s} z_{c_s} ( 
            p_s {-} \gamma_{c_s}Q_s {-} M \delta_{c_s}  {-}&\\
             \hspace{2cm} \sum_{f \in c_s}{\beta_{c_s,f} } ) {+} \sum_{c_s} \delta_{c_s} (M {+} T_s) &\nonumber\\
    
    \Psi_{L,2}(\boldsymbol{v}, \boldsymbol{\rho}, \boldsymbol{\beta}, \boldsymbol{\gamma}, \boldsymbol{\delta})  {=} 
        \sum_{s, c_s} [ \gamma_{c_s} q_{c_s} {-} \delta_{c_s} \tau_{c_s}  {+} &\\
        \hspace{2cm} \sum_{f \in c_s, \chi, j} \beta_{c_s, f} v_{c_s, f^{(j)}_{\chi}} ] {-} 
         \frac{1}{K} \sum_{f, \chi, j, k}{ \frac{\rho_{f^{(j)}_{\chi},k}}{B_k}} \,. &\nonumber
    
    % \Cc = & \sum_{s \in \Sc, c_s \in \Cc_s} \delta_{c_s} (M + T_s) \\
\end{cases}
% \end{aligned}
\end{equation}

Conveniently, the xDeSh LR problem defined above can be easily decomposed by applying primal decomposition. 
LR is indeed separable into two independent subproblems, dealing, respectively, with ($i$) the service 
configuration selection (\textbf{LR1}) and ($ii$) the xApp instantiation and resource reservation (\textbf{LR2}):
\begin{eqnarray*}
\textbf{\textsc{LR1 problem:}} \,   
    & \max_{
        \boldsymbol{z}
        }
        \enspace & \Psi_{L,1} (\boldsymbol{z}, \boldsymbol{\beta}, \boldsymbol{\gamma}, \boldsymbol{\delta})   \\
 &   \textrm{s.t.} \enspace & 
        (\ref{eq:constraint_servconf})\\
\textbf{\textsc{LR2 problem:}}\,    
 &   \max_{
        % \{ v_{c_s,f_\chi^{(j)}} , \, \rho_{f_\chi^{(j)}} \}_{
        % \forall s \in \Sc,
        % \forall c_s \in \Cc_s,
        % \forall f \in {\Fc},
        % \forall \chi \in \Xc_f,
        % \forall j}
        \boldsymbol{v}, 
        \boldsymbol{\rho}
        }
        \enspace & \Psi_{L,2} (\boldsymbol{v}, \boldsymbol{\rho}, \boldsymbol{\beta}, \boldsymbol{\gamma}, \boldsymbol{\delta})   \\
  &  \textrm{s.t.} \enspace & 
        (\ref{eq:constraint_f_notin_cs}),
        {\color{black}(\ref{eq:xApp_storage_res}),}
        (\ref{eq:constraint_budget}),
        (\ref{eq:constraint_service_disruption}),
        (\ref{eq:constraint_f_in_cs_B})\,.
\end{eqnarray*}

LR1 and LR2 are convex and therefore can be solved efficiently using standard solvers. Moreover, as LR1 and LR2 are completely independent and separable, they can be solved concurrently. 
The LR solution, also referred to as the relaxed solution, is obtained by combining the solutions of LR1 and LR2. We can then prove the following proposition:
\begin{prop}
The solution of the LR1 and LR2 problems provides a solution to the xDeSh Lagrangian Relaxation problem with an approximation ratio of $3$.
\end{prop}
\begin{proof}
LR1 is a knapsack problem with the constraint that each service can have only one active configuration at a time. 
LR2, instead, is a variant of the single-layer multi-commodity FLP, as 
 xApps correspond to facilities,  service configurations to customers, 
and services to commodities, with the additional objective of allocating 
resources to xApps to maximize the quality and minimize the services latency. 
Heuristics with known approximation ratio exist for  the knapsack ($(1{-}\epsilon)$ if the items size is within $\epsilon$ of the knapsack capacity~\cite{knapsack_approxratio})  
and the uncapacitated FLP (3 using Primal-Dual methods~\cite{FLP_approxratio}). Consequently, considering that ($i$) the relaxed solution is obtained by combining the LR1 and LR2 solutions,  ($ii$) there is only 1 possible configuration per service (i.e., $\epsilon=1$ in the knapsack), and  ($iii$)  overlooking the xApp resource allocation problem, the thesis holds.  
\end{proof}

\begin{algorithm}[tb]
\caption{Ensuring feasibility}
\label{alg:xDeSh_heuristic}
\footnotesize
\newcommand{\CommentInLine}[1]{\Statex \(\triangleright\) #1}

\nonl\textbf{Input:} $ \{ \Bar{\boldsymbol{z}} , \Bar{\boldsymbol{v}}, \Bar{\boldsymbol{\rho}} \} $ \Comment{\textit{Relaxed solution.}}

\nonl\textbf{Output:} $ \{ \Hat{\boldsymbol{z}}, \Hat{\boldsymbol{v}}, \Hat{\boldsymbol{\rho}} \} $ \Comment{\textit{Feasible solution.}}

$\Hat{\boldsymbol{z}}$ $\gets$ $\Bar{\boldsymbol{z}}$ \Comment{Accept relaxed service configuration choice.}
        
\If{ \textnormal{Eq.~(\ref{eq:constraint_f_in_cs_A}) is \textbf{not} respected for any service $s \in \Sc$} }{
    $\Hat{v}_{c_s,f_\chi^{(j)}}$ $\gets$ {\fontfamily{qcr}\selectfont xAppSelection} algorithm \Comment{Fix the relaxed xApp selection.}}

\If{\textnormal{Eq.~(\ref{eq:service_quality_constraint}) is \textbf{not} respected for any service $s \in \Sc$}}{
    $\Hat{v}_{c_s,f_\chi^{(j)}}$ $\gets$  {\fontfamily{qcr}\selectfont ServiceQualityAdjustment} \Comment{Increase the service quality by adjusting functions complexity.}
}

\If{\textnormal{Eq.~(\ref{eq:service_latency_constraint_linear}) is \textbf{not} respected for any service $s \in \Sc$}}{
    $\Hat{\rho}_{f_\chi^{(j)},\textnormal{cpu}}$ $\gets$  {\fontfamily{qcr}\selectfont ServiceLatencyAdjustment} \Comment{Reduce the service latency by adjusting xApp CPU allocation.}
}

\While{\textnormal{Eq.~(\ref{eq:constraint_budget}) or~(\ref{eq:constraint_service_disruption}) are \textbf{not} respected}}{
    
    $\tilde{s}$ $\gets$ the lowest-priority implemented service with the highest deployment cost
    
    $\Hat{z}_{c_{\tilde{s}}}$ $\gets$ 0 \Comment{Deactivate service $\tilde{s}$}

}
\end{algorithm}
\vspace{-4mm}

\subsection{Ensuring feasibility} \label{subsec:feasible_algorithm}
As mentioned, the solution of the LR1 and LR2 problems provide a solution to the xDeSh Lagrangian Relaxation problem, which however may be unfeasible.
We thus propose a multi-stage algorithm to derive a feasible solution at a later step.
As depicted in Fig.~\ref{fig:heuristic_scheme}, this algorithm receives the relaxed solution,  
and then identifies and properly rectifies any violation of the relaxed constraints (\ref{eq:service_quality_constraint}), (\ref{eq:constraint_f_in_cs_A}),  
 and~(\ref{eq:service_latency_constraint_linear}). 
The stages of this approach are reported in Alg.~\ref{alg:xDeSh_heuristic} and detailed below.

\textbf{1) Ensuring a compliant xApp selection.} The first stage assesses the compliance with constraint~(\ref{eq:service_quality_constraint}) for each chosen service configuration. 
This involves verifying if the xApps selected by the relaxed solution can meet the configuration functional requirements. 
If this condition is not met, the missing xApps need to be added to the initial solution.  %(see Line\,2 in Alg.~\ref{alg:xAppSelection_alg}). 
Then it is checked which xApps included in the solution are already deployed, and, hence, could be shared.  
Sharing is  applied only if the increase in resource consumption due to the consequent additional load would be smaller than the amount of resources needed to create a new instance of the xApp.  

\textbf{2) Meeting service quality requirements.} 
Once the xApps offering the functions of each selected service configuration have been identified, we have to ensure that the  service quality target is satisfied.
If any service fails to meet this criterion, our strategy adopts the service configuration selection offered by the relaxed solution while enhancing the complexity (output quality) of the deployed functions.

The method \textit{ServiceQualityAdjustment}
(pseudocode omitted for brevity) 
accomplishes this task by identifying the xApp that, when selected with increased complexity, contributes the most to the provided service quality at the minimum resource cost,  
i.e., it has the highest quality efficiency. 
Furthermore, after ensuring sufficient quality for all services, it is  evaluated
whether it is possible to reduce the complexity of any function to decrease resource demand 
without violating the quality constraints of any service. 
If in so doing, the quality constraints of all services are still met, then the complexity reduction is considered appropriate.

\textbf{3) Meeting service latency targets.} A similar approach is undertaken for service latency. 
Specifically, a method 
\textit{ServiceLatencyAdjustment}
(pseudocode omitted for brevity)
  increases  the CPU allocation for each xApp contributing to a service 
that fails to meet its latency target.  CPU allocation is increased first for the xApps for which the smallest CPU increase makes their latency equal to the target value. 

\textbf{4) Meeting the resource budget.} The previous stages identify  the xApps needed for the deployment of the requested services and adjust the compute resource allocation accordingly. However, the constraints (\ref{eq:constraint_budget})--(\ref{eq:constraint_service_disruption}) 
have to be fulfilled as well, i.e., it is imperative to verify the feasibility of 
the current solution and drop service requests as needed. The services to be discarded (if any)  
are those with the lowest priority and the highest deployment cost. 
This iterative procedure is repeated until the available budget is met by all resource types.

\subsection{The subgradient method} \label{subsec:subgradientmethod}
To penalize the violations of the relaxed constraints, the values of the Lagrangian multipliers 
can be determined so that the extent of such violations is minimized.
The subgradient method is a viable and computationally efficient approach to solving this~\cite{subgradientmethod}.  
The subgradient method is an iterative optimization algorithm that generalizes the gradient descent algorithm for non-differentiable functions. It involves iteratively updating the Lagrange multipliers in the direction of the subgradients of the LR problem objective function with respect to the Lagrange multipliers.  
Importantly, the subgradient method is effective with non-smooth and non-convex functions\,\cite{subgradientmethod}, as is the case of the xDeSh problem.

%%%%%%%%%%%%%%%%%%%%%%%%%%%
%%%%%%%%%%%%%%%%%%%%%%%%%%%

\section{Performance evaluation}
\label{sec:performance_eval}
We first evaluate OREO through extensive simulations  (Sec.~\ref{subsec:numerical_exp}), then we run experimental tests using a proof-of-concept implementation of the OREO framework supporting real-world RAN services and xApps (Sec.~\ref{subsec:exp_results}).

\begin{table}[tbp]
\centering
{\color{black}
\caption{Testing scenarios\vspace{-2mm}\label{tab:scenarios}}
\footnotesize
\begin{tabularx}{0.7\columnwidth}{c|Y|Y|Y}
\toprule
\textbf{Scenario} & \textbf{{$N_s$}} & {$|\Fc|$} & {$\Xc_f$} \\
\midrule
Small (\textbf{S}) & 8 & 8 & 2 \\
Medium (\textbf{M}) & 8 & 8 & 3 \\
Large (\textbf{L}) & 10 & 8 & 3 \\
Extra Large (\textbf{XL}) & 12 & 10 & 3 \\
\bottomrule
\end{tabularx}
}
\vspace{-2mm}
\end{table}

\begin{figure}[tbp]
    \centering
    \includegraphics[width=0.36\textwidth]{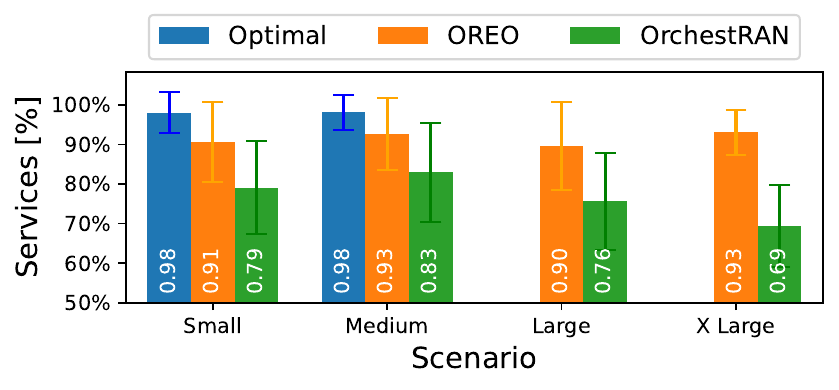}
    \vspace{-2mm}
    \hspace{2mm}
    \includegraphics[width=0.36\textwidth]{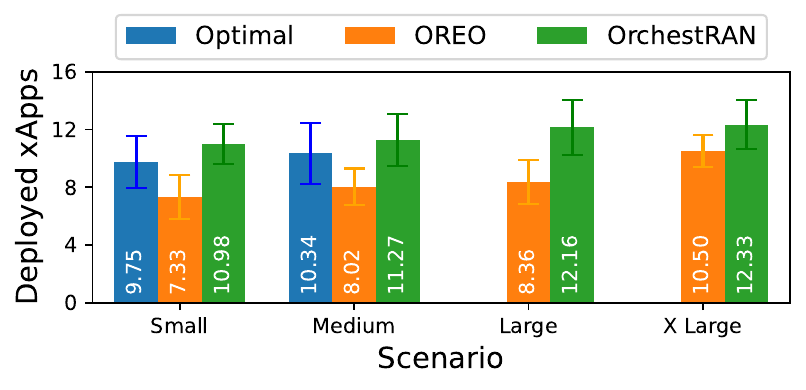}
 \caption{Numerical results: Percentage of services (top) and  xApps (bottom) deployed by Optimal, OREO, and OrchestRAN.}
    \label{fig:numericalresults_services}
\vspace{-6mm} 
\end{figure}

\begin{figure*}[!tbph]
\centering
    {\includegraphics[width=.32\textwidth]{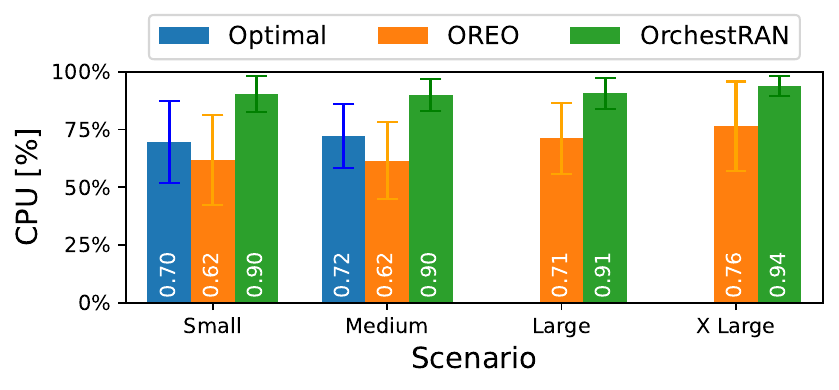}}
    {\includegraphics[width=.32\textwidth]{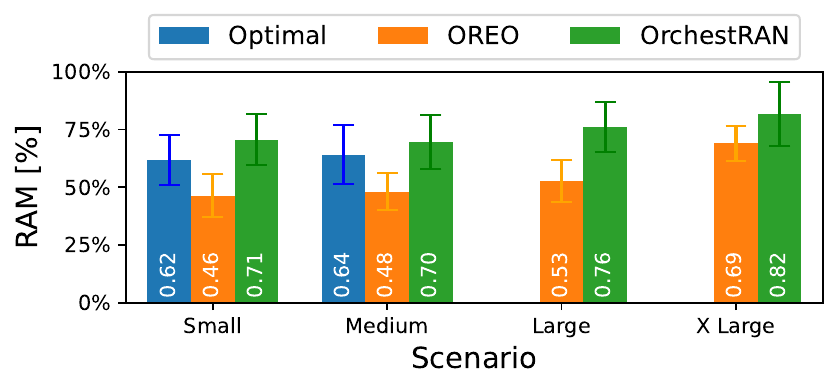}}
    {\includegraphics[width=.32\textwidth]{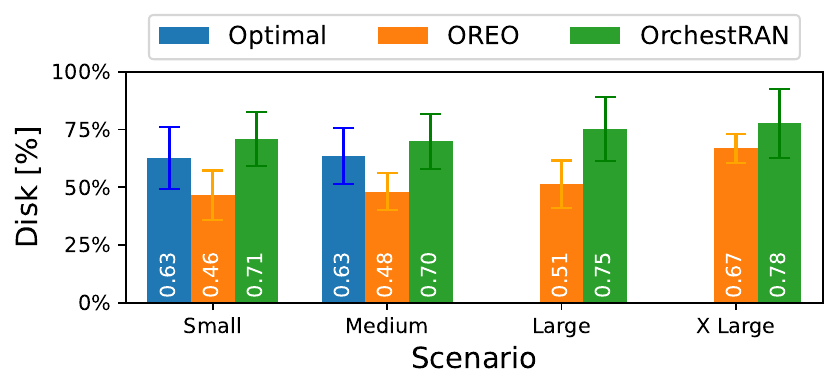}}
\vspace*{-4mm}
\caption{Numerical results: CPU (left), RAM (center), and Disk (right) resources used by  Optimal, OREO, and OrchestRAN.
    \label{fig:numericalresults_resources}}
\vspace{-4mm}
\end{figure*}

\subsection{Numerical analysis}
\label{subsec:numerical_exp}

To test the effectiveness of OREO at scale, we developed a Python simulator. In the considered scenario, an MNO generates requests for a set of $N_s$ services. The MNO requests are forwarded to the non-RT RIC, where the OREO selects the most amenable configuration among 3 possible ones. Each configuration involves at most 4 RAN functions among the $|\Fc|$ available, with each function featuring $\Xc_f$ different levels of complexity. 
We consider {\color{black}4} scenarios of different scales as illustrated in Tab.\,\ref{tab:scenarios} and present average results over 100 runs.

\textbf{Benchmarks.}
We compare OREO against two alternative policies: the ``Optimal'' policy, which uses Gurobi  
% \footnote{https://support.gurobi.com/hc/en-us}
to optimally solve the xDeSh problem, and OrchestRAN~\cite{orchestran}, a state-of-the-art O-RAN orchestrator. 
{\color{black} 
For fairness, since OrchestRAN does not consider services as compositions of interconnected xApps, we let it handle
every possible combination of ($i$) each configuration of a service requested by the MNO and ($ii$) every available complexity factor for the involved functions, as distinct xApps.
}

\begin{table}[b]
{\color{black}
\centering
\vspace{-6mm}
\caption{Empirical approximation ratio and run times\vspace{-2mm}\label{tab:approximationratio}}
\footnotesize
    \begin{tabularx}{1.\columnwidth}{c|YYYYY|YYY|Y}
\toprule
\multicolumn{1}{c}{} & \multicolumn{5}{c}{\textbf{OREO}} & \multicolumn{3}{c}{\textbf{OrchestRAN}} & \multicolumn{1}{c}{\textbf{Opt.}} \\
\cmidrule(rl){2-6} \cmidrule(rl){7-9} \cmidrule(rl){10-10}
\textbf{} & $\alpha$ & $0.9\alpha$ &  $\Bar{\alpha}$ &  $t$ &  $S$ & $\alpha$ & $0.9\alpha$ &  $\Bar{\alpha}$ &  $t$ \\
\midrule
\textbf{S} & 0.77 & 0.86 & 0.88 & 3.44 & 15.75 & 0.23 & 0.35 & 0.38 & 54.18 \\
\textbf{M} & 0.81 & 0.86 & 0.89 & 4.51 & 41.33 & 0.16 & 0.34 & 0.37 & 186.40 \\
\textbf{L} & $-$ & $-$ & $-$ & 6.33 & $-$ & $-$ & $-$ & $-$ & $-$ \\
\textbf{XL} & $-$ & $-$ & $-$ & 10.31 & $-$ & $-$ & $-$ & $-$ & $-$ \\
\bottomrule
\end{tabularx}
}
\end{table}

\begin{figure}[tbp]
    \centering
    \includegraphics[width=0.36\textwidth]{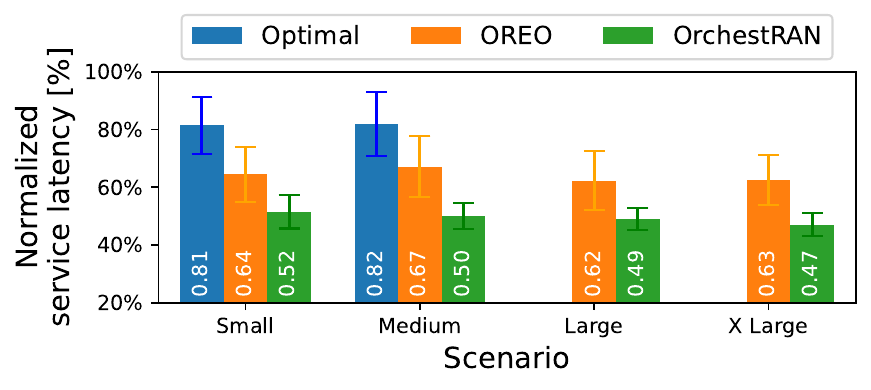}
    \vspace*{-4mm} 
    \caption{Normalized latency performance of RAN services offered by  Optimal, OREO, and OrchestRAN.}
    \label{fig:numericalresults_latency}
\vspace{-4mm} 
\end{figure}

\textbf{Approximation ratio estimate and run times.}
Tab.~\ref{tab:approximationratio} reports the numerical estimates of the approximation ratio $\alpha$ (i.e., the ratio of the heuristic performance to the optimum) and the execution times for the implemented orchestration policies. Specifically, the table presents the 90\% confidence interval lower-bound ($0.9\alpha$) and the average value $\Bar{\alpha}$, for both OREO and OrchestRAN.
Note that, for the two largest scenarios {\color{black}(i.e., L and XL)}, where the number of optimization variables is over $10^4$, we were unable to compute such metrics, as computing the optimum becomes impractical.  

OREO consistently provides solutions that are within {\color{black}0.75} from the optimum across all tested scenarios. With 90\% confidence, the estimated ratio increases to 0.86; also, the relatively narrow confidence interval suggests that OREO's performance consistently remains close to the average values.  
In contrast, OrchestRAN's performance gap widens with scenario size, and even in the simplest scenario (S), it achieves an approximation ratio that is {\color{black}70\%} worse than that of OREO.

\begin{figure*}[!tbph]
\centering
    {\includegraphics[width=.32\textwidth]{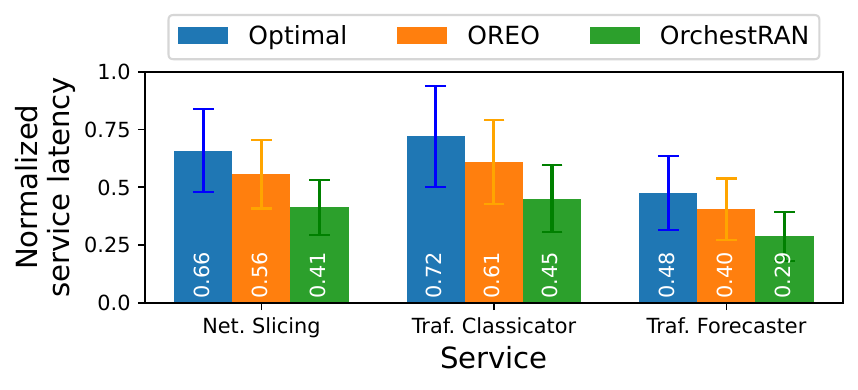}}
    {\includegraphics[width=.32\textwidth]{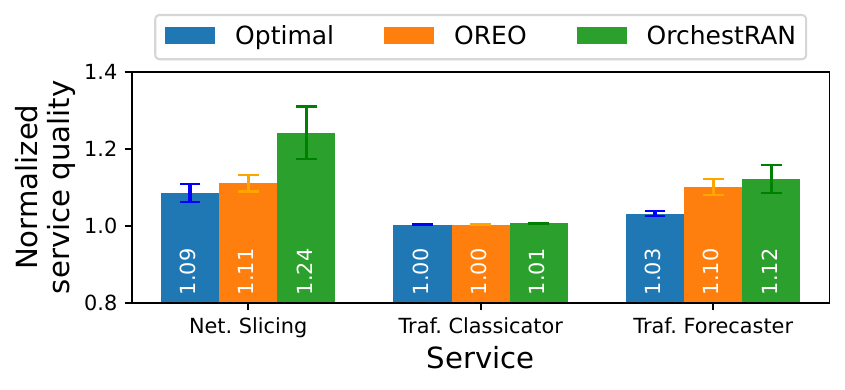}}{\includegraphics[width=.31\textwidth,raise=3.5mm]{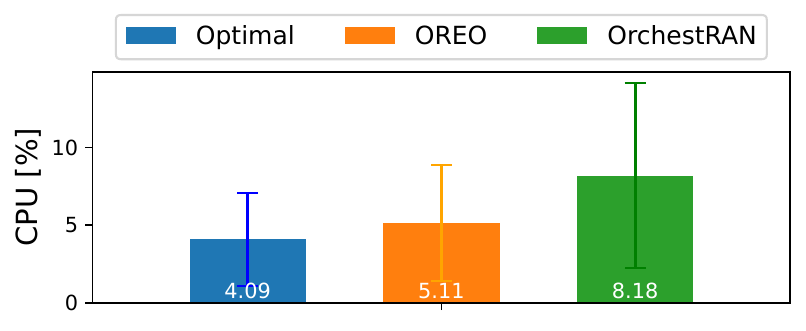}}
\vspace*{-2mm}
\caption{Testbed results: normalized service latency (left) and quality (center), and CPU consumption (right) of OREO, Optimal and OrchestRAN.
    \label{fig:experimental_results}
}
\vspace{-4mm}
\end{figure*}

Regarding execution times, OREO is highly efficient with averages below 10 seconds, even for large-scale scenarios, proving to be 
% from 3 to 40 times faster
{\color{black}up to 40 times faster}
than the optimum.

\textbf{Deployed services and xApps.}
Examining the percentage of services provided by OREO and its counterparts (Fig.~\ref{fig:numericalresults_services} (top)), the Optimal provides approximately 95\% deployed services
% as the scenario varies from T to L,
{\color{black}within the two smallest scenarios (i.e., S and M).}
{\color{black}This implies that the tested scenarios are demanding w.r.t. the near-RT RIC capabilities.} 
OREO achieves performance comparable (within 8\%) to the optimum.
On the contrary, OrchestRAN results in a significant decrease in the service implementation rate, as it yields 
% 15\%
{\color{black}19\%}
fewer services compared to the optimum.
The underlying reasons can be found in Fig.~\ref{fig:numericalresults_services} (bottom), which illustrates the average number of xApps deployed by OREO and its alternatives in the considered scenarios. Although OREO deploys more services than OrchestRAN, it implements fewer xApps, thanks to its superior ability to share xApps and adjust the resources allocated to them. In particular, OREO instantiates
% 28.2\%
{\color{black}30.7\%} and
20.7\% fewer xApps than, respectively, OrchestRAN and the optimum, which however can better use the xApps it deploys to ultimately offer more services.

\textbf{Near-RT RIC resource consumption.}
The higher ability to share xApps exhibited by OREO in comparison to its benchmark is confirmed by its lower resource utilization at the near-RT RIC. While OrchestRAN employs almost the whole CPU budget under all scenarios (Fig.~\ref{fig:numericalresults_resources} (left)) and takes a significant toll also on RAM (Fig.~\ref{fig:numericalresults_resources} (center)) and Disk (Fig.~\ref{fig:numericalresults_resources} (right)), OREO yields substantial resource savings 
% (25.6\% of CPU).
{\color{black}(31.1\% of CPU).}
Notice that also Optimal uses a significant amount of resources, as minimizing resource consumption is not the only objective of the xDeSh problem. Accordingly, relatively to OREO, it makes up for the larger resource consumption with a higher number of deployed services.

\textbf{Meeting service requirements.} OREO properly scales the resources allocated to the xApps according to the aggregated load of the services sharing the xApp, so that it can successfully fulfill  the service requirements. Focusing on service latency for brevity, Fig.~\ref{fig:numericalresults_latency} shows the normalized service latency (i.e., the ratio of actual to target service latency) over the test scenarios.
Both OREO and its counterparts can meet the target service latency in all cases, {\color{black} as their
 average normalized service latency always remains below 1.} The Optimal and OREO policies indeed align xApps processing latency with the service having the most stringent requirements among those sharing the xApp. Conversely,  OrchestRAN overlooks the scaling issue related to computing resources, resulting in sub-optimal decisions.
Moreover, looking at Fig.~\ref{fig:numericalresults_latency} and Fig.~\ref{fig:numericalresults_resources} together, it is evident that OREO, similarly to the Optimal policy, fulfills service requirements while saving substantial resources with respect to OrchestRAN.

\subsection{Testbed setup and results}
\label{subsec:exp_results}

%{\bf Testbed design and setup.}
Our testbed integrates the OREO architecture (Fig.~\ref{fig:oreo_system_design}) with an emulated softwarized cellular base station. Our experimental platform includes an O-Cloud environment where xApps can be instantiated as docker containers.  

In our experiments, an MNO module generates a service request every 100\,s. Each request involves from 1 to 3 services, chosen with equal probability among those listed below, and with a randomly-selected target latency-quality pair. 
The possible latency targets for all services are 0.1, 0.2, and 0.5\,s; the service quality targets, instead, are  service-specific.

\textbullet\,{\em Traffic forecasting:} it predicts user traffic loads, as required by many  proactive RAN controllers.
It consists of a single configuration with a traffic forecaster xApp $f_1$ that uses a Long Short-Term Memory (LSTM) model. The xApp supports 3 complexity levels, corresponding to different numbers of input samples (6 to 30) and LSTM layers (1 to 4), and different sizes of the LSTM hidden layers (1 to 5). The possible service quality targets are 0.9, 0.925, or 0.95.

\textbullet\,{\em Traffic classification:} it  identifies the applications generating monitored traffic flows, e.g., so that application-dedicated policies can be applied.
The service includes a single configuration with a traffic classificator xApp $f_2$ that uses a Random Forest classifier to label traffic samples. The xApp supports 3 complexity levels, depending on the number of  applications that can generate traffic flows (3 to 20). The service can be requested with target quality values equal to 0.7, 0.8, or 0.9.

\textbullet\,{\em Network slicing:} it optimizes the allocation of radio resources to network slices (eMBB, uRLLC, or mMTC). The service can be implemented with 4 different configurations: $f_1 {+} f_2 {\rightarrow} f_3$; $f_1{\rightarrow}f_3$; $f_2{\rightarrow} f_3$; $f_3$, where $f_3$ is the RL-based slicing policy introduced in \cite{xApp_ColO-RAN}. The service can be requested  with a target quality value of 0.6, 0.8, or 0.9.

Once received, the non-RT RIC processes the service requests and uses OREO (or one of the benchmarks) to decide the set of xApps and their resource allocation that best provide the service. Then, the xApp lifecycle manager in the non-RT RIC instructs the instantiation of the xApps in the near-RT RIC accordingly. Each service has a 100-s lifetime and the traffic scenario is  determined using the  dataset in \cite{xApp_ColO-RAN},  which refers to a base station serving 6 users, evenly split into eMBB, URLLC, and mMTC-like traffic patterns. 

{\bf Results.}
Fig.~\ref{fig:experimental_results} presents the average value and the 90\% confidence interval of the normalized (i.e., the ratio of actual to target) service latency (left), service quality (center), and CPU consumption (right).  
Different than some of the scenarios presented in the previous subsection,  the reduced scale of the scenario considered here allows both OREO and its benchmarks to meet all the service requirements.
However, OREO selects better service configurations, similar to the Optimal policy, which renders a reduced utilization of CPU resources when compared to OrchestRAN. 
In fact, OrchestRAN struggles to find the right balance between  service requirements and resource utilization. Specifically, although OrchestRAN provides lower service latency and higher service quality, it uses 60\% and 100\% higher CPU consumption than OREO and Optimal, respectively. The savings attained by OREO, only consuming 24.9\% more CPU resources than Optimal, are achieved by pushing the latency (23\%) and quality (5\%) requirements  closer to their target than OrchestRAN.

\section{Related Work}
\label{sec:related_work}
Recently, considerable attention has been paid to the design of near-RT RIC xApps~\cite{ORANWG3_NearRTArch}, aiming at optimally controlling O-RAN networks. Relevant examples are xApps for network traffic classification~\cite{xApp_TraffClass1,xApp_TraffClass2} and network load forecasting~\cite{xApp_TraffForec1,xApp_TraffForec2}. Further, an xApp can lay down policies to define RAN slices, as done in~\cite{xApp_ColO-RAN,xAPP_NetSlic1,xAPP_NetSlic2}, or to manage the RAN radio resources~\cite{CAREM,xApp_RRM1}.

In this context, proficiently managing and orchestrating the RAN becomes even more crucial, necessitating the best exploitation of the multitude of multi-vendor solutions~\cite{ran_orchest}. Despite the large body of work existing on RAN orchestration~\cite{vrain}, few studies have tackled network intelligence management in O-RAN.
Among these,~\cite{energy_orch_vran} proposes a computationally efficient orchestrator for energy consumption optimization in virtual RANs.
The study in~\cite{orch_inference} introduces, instead, a distributed dynamic policy for instantiating inference models, providing performance guarantees.

The closest work to ours is  OrchestRAN~\cite{orchestran}, a pioneering O-RAN orchestrator that identifies the optimal set of xApps to deploy and their location to offer the services requested by network operators while meeting the performance targets. It assumes the RAN services to be monolithic, i.e., solely provided by an xApp, thus limiting the possibility of sharing low-level operations between services. Thus, \cite{orchestran}, unlike our work, disregards the issue of RAN resources consumption --  a major contribution to MNOs' OPEX~\cite{energy_orch_vran}.

\section{Conclusions}
\label{sec:concl}

We proposed OREO, an O-RAN orchestrator  for xApp-based network services. Unlike previous works, OREO considers that services can be deployed through different sets of  shareable elementary functions, with each function possibly yielding an output of different quality and implemented as an xApp. To limit resource consumption while fulfilling the service requirements, OREO properly tunes xApps at different complexity factors, which correspond to different quality-latency-resource demand tradeoffs.  Numerical results show that OREO performs close to the optimum, and, compared to the state of the art, it allocates  more  (16.2\% on average and 35\% in the largest scenario) services and consumes  less CPU (25.6\% on average and over 31\% in small-medium scenarios), while meeting the service requirements. Experimental results obtained through our proof-of-concept testbed confirm the good performance of OREO and its ability to efficiently deploy xApps, yielding a 37.5\% reduction on CPU consumption with respect to the state of the art.

\section*{Acknowledgments}
This work was supported by the European Commission through Grant No.\,101095890 (PREDICT-6G project), Grant No.\,101096379 (CENTRIC project), and Grant No.\,101017109 (DAEMON project), and by the EU under the Italian NRRP  of NextGenerationEU, partnership on “Telecommunications of the Future” (PE0000001 - program “RESTART”).
%\newpage\clearpage

\bibliographystyle{IEEEtran}
\bibliography{refs}

\begin{thebibliography}{10}
\providecommand{\url}[1]{#1}
\csname url@samestyle\endcsname
\providecommand{\newblock}{\relax}
\providecommand{\bibinfo}[2]{#2}
\providecommand{\BIBentrySTDinterwordspacing}{\spaceskip=0pt\relax}
\providecommand{\BIBentryALTinterwordstretchfactor}{4}
\providecommand{\BIBentryALTinterwordspacing}{\spaceskip=\fontdimen2\font plus
\BIBentryALTinterwordstretchfactor\fontdimen3\font minus \fontdimen4\font\relax}
\providecommand{\BIBforeignlanguage}[2]{{%
\expandafter\ifx\csname l@#1\endcsname\relax
\typeout{** WARNING: IEEEtran.bst: No hyphenation pattern has been}%
\typeout{** loaded for the language `#1'. Using the pattern for}%
\typeout{** the default language instead.}%
\else
\language=\csname l@#1\endcsname
\fi
#2}}
\providecommand{\BIBdecl}{\relax}
\BIBdecl

\bibitem{6g_network}
Z.~Zhang, Y.~Xiao, Z.~Ma, M.~Xiao, Z.~Ding, X.~Lei, G.~K. Karagiannidis, and P.~Fan, ``{6G Wireless Networks: Vision, Requirements, Architecture, and Key Technologies},'' \emph{IEEE Vehicular Technology Magazine}, vol.~14, no.~3, pp. 28--41, 2019.

\bibitem{ORANWG1_Arch}
{O-RAN Working Group 1}, ``{O-RAN Architecture Description 9.00},'' June 2023.

\bibitem{understanding_oran}
M.~Polese, L.~Bonati, S.~D’Oro, S.~Basagni, and T.~Melodia, ``{Understanding O-RAN: Architecture, Interfaces, Algorithms, Security, and Research Challenges},'' \emph{IEEE Communications Surveys \& Tutorials}, vol.~25, no.~2, pp. 1376--1411, 2023.

\bibitem{oran_disrupting_vran_ecosystem}
A.~Garcia-Saavedra and X.~Costa-Pérez, ``{O-RAN: Disrupting the Virtualized RAN Ecosystem},'' \emph{IEEE Communications Standards Magazine}, vol.~5, no.~4, pp. 96--103, 2021.

\bibitem{ORANWG1_UseCases}
{O-RAN Working Group 1}, ``{Use Cases Analysis Report 11.0},'' June 2023.

\bibitem{RAN_related_costs}
K.~Johansson, A.~Furuskar, P.~Karlsson, and J.~Zander, ``{Relation between base station characteristics and cost structure in cellular systems},'' in \emph{{2004 IEEE 15th International Symposium on Personal, Indoor and Mobile Radio Communications (IEEE Cat. No.04TH8754)}}, vol.~4, 2004, pp. 2627--2631 Vol.4.

\bibitem{orchestration_requirements}
M.~Camelo, L.~Cominardi, M.~Gramaglia, M.~Fiore, A.~Garcia-Saavedra, L.~Fuentes, D.~De~Vleeschauwer, P.~Soto-Arenas, N.~Slamnik-Krijestorac, J.~Ballesteros, C.-Y. Chang, G.~Baldoni, J.~M. Marquez-Barja, P.~Hellinckx, and S.~Latré, ``{Requirements and Specifications for the Orchestration of Network Intelligence in 6G},'' in \emph{{2022 IEEE 19th Annual Consumer Communications \& Networking Conference (CCNC)}}, 2022, pp. 1--9.

\bibitem{orchestran}
S.~D’Oro, L.~Bonati, M.~Polese, and T.~Melodia, ``{OrchestRAN: Network Automation through Orchestrated Intelligence in the Open RAN},'' in \emph{{2022 IEEE Conference on Computer Communications (INFOCOM)}}, 2022, pp. 270--279.

\bibitem{service_function_chaining}
J.~Zhang, Z.~Wang, N.~Ma, T.~Huang, and Y.~Liu, ``{Enabling efficient service function chaining by integrating NFV and SDN: Architecture, challenges and opportunities},'' \emph{IEEE Network}, vol.~32, no.~6, pp. 152--159, 2018.

\bibitem{xApp_ColO-RAN}
M.~Polese, L.~Bonati, S.~D'Oro, S.~Basagni, and T.~Melodia, ``{ColO-RAN: Developing Machine Learning-based xApps for Open RAN Closed-loop Control on Programmable Experimental Platforms},'' \emph{IEEE Transactions on Mobile Computing}, pp. 1--14, 2022.

\bibitem{o-cloud}
{O-RAN Working Group 6}, ``{Cloud Platform Reference Designs 2.0},'' February 2021.

\bibitem{Cohen}
I.~Cohen, C.~F. Chiasserini, P.~Giaccone, and G.~Scalosub, ``{Dynamic Service Provisioning in the Edge-cloud Continuum with Provable Guarantees},'' \emph{arXiv preprint arXiv:2202.08903}, 2022.

\bibitem{mm1_1}
F.~Ben~Jemaa, G.~Pujolle, and M.~Pariente, ``{QoS-Aware VNF Placement Optimization in Edge-Central Carrier Cloud Architecture},'' in \emph{{2016 IEEE Global Communications Conference (GLOBECOM)}}, 2016, pp. 1--7.

\bibitem{mm1_2}
R.~Gouareb, V.~Friderikos, and A.-H. Aghvami, ``{Virtual Network Functions Routing and Placement for Edge Cloud Latency Minimization},'' \emph{IEEE Journal on Selected Areas in Communications}, vol.~36, no.~10, pp. 2346--2357, 2018.

\bibitem{lagr_heuristic_2CFLP_1}
A.~Klose, ``{A Lagrangean relax-and-cut approach for the two-stage capacitated facility location problem},'' \emph{European journal of operational research}, vol. 126, no.~2, pp. 408--421, 2000.

\bibitem{lagr_heuristic_2CFLP_2}
------, ``{An LP-Based Heuristic for Two-Stage Capacitated Facility Location Problems},'' \emph{Journal of the Operational Research Society}, vol.~50, no.~2, pp. 157--166, 1999.

\bibitem{lagr_heuristic_2UFLP}
B.~Gendron, P.-V. Khuong, and F.~Semet, ``{A Lagrangian-Based Branch-and-Bound Algorithm for the Two-Level Uncapacitated Facility Location Problem with Single-Assignment Constraints},'' \emph{Transportation Science}, vol.~50, no.~4, pp. 1286--1299, 2016.

\bibitem{knapsack_approxratio}
A.~M. Frieze, M.~R. Clarke \emph{et~al.}, ``{Approximation Algorithms for the m-Dimensional 0-1 Knapsack Problem: Worst-Case and Probabilistic Analyses},'' \emph{European Journal of Operational Research}, vol.~15, no.~1, pp. 100--109, 1984.

\bibitem{FLP_approxratio}
K.~Jain and V.~V. Vazirani, ``{Approximation Algorithms for Metric Facility Location and k-Median Problems Using the Primal-Dual Schema and Lagrangian relaxation},'' \emph{Journal of the ACM (JACM)}, vol.~48, no.~2, pp. 274--296, 2001.

\bibitem{subgradientmethod}
M.~Held, P.~Wolfe, and H.~P. Crowder, ``{Validation of Subgradient Optimization},'' \emph{Mathematical programming}, vol.~6, pp. 62--88, 1974.

\bibitem{ORANWG3_NearRTArch}
{O-RAN Working Group 3}, ``{O-RAN Near-RT RAN Intelligent Controller Near-RT RIC Architecture 2.01},'' March 2022.

\bibitem{xApp_TraffClass1}
J.~Shi, M.~Liu, C.~Hou, M.~Jiang, and G.~Xiong, ``{Online Encrypted Mobile Application Traffic Classification at the Early Stage: Challenges, Evaluation Criteria, Comparison Methods},'' in \emph{{2021 IEEE 6th International Conference on Computer and Communication Systems (ICCCS)}}, 2021, pp. 1128--1135.

\bibitem{xApp_TraffClass2}
Y.~Li, B.~Liang, and A.~Tizghadam, ``{Robust Online Learning against Malicious Manipulation and Feedback Delay With Application to Network Flow Classification},'' \emph{IEEE Journal on Selected Areas in Communications}, vol.~39, no.~8, pp. 2648--2663, 2021.

\bibitem{xApp_TraffForec1}
A.~M. Nagib, H.~Abou-Zeid, H.~S. Hassanein, A.~Bin~Sediq, and G.~Boudreau, ``{Deep Learning-Based Forecasting of Cellular Network Utilization at Millisecond Resolutions},'' in \emph{{ICC 2021 - IEEE International Conference on Communications}}, 2021, pp. 1--6.

\bibitem{xApp_TraffForec2}
N.~Salhab, R.~Langar, R.~Rahim, S.~Cherrier, and A.~Outtagarts, ``{Autonomous Network Slicing Prototype Using Machine-Learning-Based Forecasting for Radio Resources},'' \emph{IEEE Communications Magazine}, vol.~59, no.~6, pp. 73--79, 2021.

\bibitem{xAPP_NetSlic1}
S.~Bakri, P.~A. Frangoudis, A.~Ksentini, and M.~Bouaziz, ``{Data-Driven RAN Slicing Mechanisms for 5G and Beyond},'' \emph{{IEEE Transactions on Network and Service Management}}, vol.~18, no.~4, pp. 4654--4668, 2021.

\bibitem{xAPP_NetSlic2}
J.~Mei, X.~Wang, K.~Zheng, G.~Boudreau, A.~B. Sediq, and H.~Abou-Zeid, ``{Intelligent Radio Access Network Slicing for Service Provisioning in 6G: A Hierarchical Deep Reinforcement Learning Approach},'' \emph{IEEE Transactions on Communications}, vol.~69, no.~9, pp. 6063--6078, 2021.

\bibitem{CAREM}
S.~Tripathi, C.~Puligheddu, C.~F. Chiasserini, and F.~Mungari, ``{A Context-Aware Radio Resource Management in Heterogeneous Virtual RANs},'' \emph{IEEE Transactions on Cognitive Communications and Networking}, vol.~8, no.~1, pp. 321--334, 2022.

\bibitem{xApp_RRM1}
S.~Shen, T.~Zhang, S.~Mao, and G.-K. Chang, ``{DRL-Based Channel and Latency Aware Radio Resource Allocation for 5G Service-Oriented RoF-MmWave RAN},'' \emph{Journal of Lightwave Technology}, vol.~39, no.~18, pp. 5706--5714, 2021.

\bibitem{ran_orchest}
A.~Arnaz, J.~Lipman, M.~Abolhasan, and M.~Hiltunen, ``{Towards integrating intelligence and programmability in open radio access networks: A comprehensive survey},'' \emph{Ieee Access}, 2022.

\bibitem{vrain}
J.~A. Ayala-Romero, A.~Garcia-Saavedra, M.~Gramaglia, X.~Costa-Pérez, A.~Banchs, and J.~J. Alcaraz, ``{vrAIn: Deep Learning Based Orchestration for Computing and Radio Resources in vRANs},'' \emph{IEEE Transactions on Mobile Computing}, vol.~21, no.~7, pp. 2652--2670, 2022.

\bibitem{energy_orch_vran}
R.~Singh, C.~Hasan, X.~Foukas, M.~Fiore, M.~K. Marina, and Y.~Wang, ``{Energy-Efficient Orchestration of Metro-Scale 5G Radio Access Networks},'' in \emph{{2021 IEEE Conference on Computer Communications (INFOCOM)}}, 2021, pp. 1--10.

\bibitem{orch_inference}
T.~Si~Salem, G.~Castellano, G.~Neglia, F.~Pianese, and A.~Araldo, ``{Towards Inference Delivery Networks: Distributing Machine Learning with Optimality Guarantees},'' in \emph{{2021 19th Mediterranean Communication and Computer Networking Conference (MedComNet)}}, 2021, pp. 1--8.

\end{thebibliography}

\end{document}